\documentclass[12pt]{iopart}

\usepackage{iopams}
\usepackage{graphicx}
\usepackage{bm}
\usepackage{amsfonts}
\newcommand{\ket}[1]{\mbox{$\mid \! #1 \, \rangle$}}
\newcommand{\bra}[1]{\mbox{$\langle \, #1 \! \mid$}}

\newcommand{\eins}{\mbox{$1 \hspace{-1.0mm}  {\bf l}$}}

\newcommand{\WW}{\ensuremath{\mathcal{W}}}

\begin{document}

\title[]{Experimental high fidelity six-photon entangled state for telecloning protocol}

\author{Magnus R{\aa}dmark$^{1}$, Marek \.{Zukowski}$^{2}$ and Mohamed Bourennane$^{1}$}

\address{$^{1}$Physics Department, Stockholm University, SE-10691 Stockholm, Sweden\\
$^{2}$Institute for Theoretical Physics and Astrophysics,
Uniwersytet Gda\'{n}ski, PL-80-952 Gda\'{n}sk, Poland}
\ead{radmark@physto.se}
\begin{abstract}
We experimentally generate and characterize a six-photon
polarization entangled state, which is usually called
``$\Psi_6^{+}$''. This is realized with a  filtering procedure of
triple emissions of entangled photon pairs from a single source,
which does not use any interferometric overlaps. The setup is very
stable and we observe the six-photon state with high fidelity. The
observed state can be used for demonstrations of telecloning and
secret sharing protocols.
\end{abstract}

\pacs{03.67.-a, 03.67.Hk, 03.65.Ud, 42.65.Lm}

\maketitle

\section{Introduction}
Multiphoton interference is a rich source of non-classical effects.
As there exist sources that directly produce entangled states of pairs of photons,
in the first stage of the development of the field,
experiments concentrated on two-photon interference~\cite{CLAUSER,ASPECT,ALLEY,MANDEL}. 
With the Greenberger-Horne-Zeilinger~\cite{GHZ} paper it became evident that if one goes towards three- or more photon interference effects, a new and extremely rich realm of ultra non-classical phenomena can be discovered. The emergence of a new field of physics and technology, quantum information, and particularly quantum communication and cryptography~\cite{BB84,EKERT,GISIN} transformed such phenomena into a new playground of applied physics. This interplay between new photonic processes and their information applications continues and accelerates. The teleportation~\cite{TELEPORTATION} experiment involving two separate spontaneous emissions of entangled pairs clearly demonstrated, that three-photon effects are potentially observable in the laboratory, and demonstrated a basic quantum informational process requiring {\em three-particle interference}~\cite{TELEPORTATION-EXP}. 
Multiphoton interference (by which we understand three- or more photon effects) has since then been used in many experiments for testing the foundations of quantum mechanics, including the generation of GHZ correlations~\cite{GHZ3,GHZ4,EGBKZW03},
and in demonstrations of basic quantum information protocols~\cite{review}. A summary of these efforts can be found in e.g.~\cite{BEZ_book,Kok_LinOpt}.
In contradistinction to entangled pairs, multiphoton effects require state engineering, since the only way we obtain them is by
utilizing two or more entangled pair generations in several sources, or via multiple emissions in one source, and suitable measurement procedures which swap~\cite{ZZHE}, or process~\cite{ZHWZ97} entanglement. It requires special techniques~\cite{ZZW95,RARITY}, which are being continuously improved (for recent advances see e.g. \cite{BANASZEK}); these include new schemes~\cite{ZW01} and sources.  This progress now allows observations of six-photon interference processes with reasonable count rates. The trailblazing paper was in this case the one by Lu {\it et al.}~\cite{LZGGZYGYP07}. Thereafter, six-photon entanglement effects were reported in various experimental configurations~\cite{ANTON,HARALD,MOHAMED1,MOHAMED2}. As this type of effects are now under our control, one could now advance to multiparty communication protocols which require sixpartite entanglement.


An example of such a protocol is telecloning, where, in order to produce three imperfect copies 
of a qubit state, one requires a specific six-qubit entangled state usually called $\Psi_{6}^{+}$.
In this protocol a sender (Alice) wishes, via quantum channels, to
distribute {\em quantum} information, e.g. the state of an unknown qubit $\ket{X}$, to
several partners placed at different remote locations. The no-cloning
theorem forbids her to copy or to broadcast totally unknown quantum
information~\cite{WZ82}. Fortunately laws of quantum physics allow
Alice to transmit the state to her associates with a significant fidelity
up to $F = (2M+1)/3M$, where $M$ is the number of receiving parties~\cite{BH96,BDEFMS98}. 
The Murao {\it et al.}~\cite{MJPV99} `telecloning' scheme allows her to perform an optimal
broadcasting of quantum information to three partners. In this
protocol, Alice and her partners must initially share $\ket{\Psi_{6}^{+}}$. Alice should have three qubits (two serve as passive ancillas) from $\ket{\Psi_{6}^{+}}$ in addition to the qubit $\ket{X}$,
while her partners should have one qubit each. Alice then performs a local joint
(Bell) measurement on the unknown qubit and one of her qubits from
$\ket{\Psi_{6}^{+}}$. Finally she sends a classical two-bit message
to her three partners, informing them of her measurement result, and they 
perform local unitary transformations on their qubits according to
Alice's message. The final quantum states of each of Alice's
partners are now optimal copies of her initial state with the maximal possible
fidelity, $F = 7/9$. The telecloning protocol combines an optimal quantum cloning
machine and the teleportation protocol. 
The full experimental implementation of telecloning requires seven-photon interference, but here the aim was to generate the specific six-photon state $\Psi_{6}^{+}$ which is required for three-location-telecloning.
 It has also been shown theoretically that $\ket{\Psi_{6}^{+}}$ can be used for secure quantum multiparty
cryptographic protocols, such as the  six-party secret sharing
protocol~\cite{HBB99,GKBW07}.


We report the first experimental generation of $\ket{\Psi_{6}^{+}}$. In strong contrast to the first six-photon entanglement experiment~\cite{LZGGZYGYP07}, in which a generalization of the overlap schemes suggested in~\cite{ZHWZ97}
was used, we here achieve six-photon entanglement by pulse pumping just {\em one crystal}, extracting the third order processes, and distributing the photons into six spatial modes. 
The required indistinguishability of photons is obtained by the now standard techniques employing suitable filtering~\cite{ZZW95,RARITY}. 
That is, we generalize the procedure theoretically proposed
in~\cite{ZW01}, which has been used to produce the four-qubit singlet 
state $\Psi_{4}^{-}$~\cite{GBEKW03}. This method was tested in, e.g.~\cite{MOHAMED1} 
and~\cite{MOHAMED2} (for related experiments see~\cite{HARALD} and~\cite{ANTON}). As there are no
interferometric overlaps in the setup, it is very stable. 

The paper is organized as follows.
In section \ref{sec:setup} we will describe our experimental setup. In section \ref{sec:results} we will show our measurement results and we calculate the quantum correlations of the state.
Further on, we will show its robustness against photon loss and describe how we detect sixpartite entanglement in the state.
Finally we will give a conclusion in section \ref{sec:conclusion}.


\section{Experimental setup}\label{sec:setup} Let us start with a
brief explanation  of the theory of the used Parametric
Down-Conversion (PDC) process and then a detailed description of our
experimental setup~\cite{MOHAMED1}.

The state of two phase matched modes of the multiphoton emission
that results out of a single pulse acting on a type-II PDC crystal
is given by
\begin{equation}
C \mathrm{exp}(-i\alpha(a_{0H}^{\dagger}b_{0V}^{\dagger}+
a_{0V}^{\dagger}b_{0H}^{\dagger}))\ket{0}, 
\label{emission}
\end{equation}
where $a_{0H}^{\dagger}$ ($b_{0V}^{\dagger}$) is the creation
operator for one horizontal (vertical) photon in mode $a_{0}$
($b_{0}$), and conversely; $C=1/\sqrt{\sum_{n=0}^\infty(1+n)|\alpha|^{2n}}$ is a normalization constant, $\alpha$
is a function of pump power, non-linearity and length of the crystal
and $\ket{0}$ denotes the vacuum state. This is a good description
of the state, provided one collects the photons under conditions
that allow the indistinguishability between separate two-photon
emissions~\cite{ZZW95}. The third order term in the
 expansion of (\ref{emission}), corresponds to the
emission of six photons. In our experiments these photons are
distributed into six modes using $50-50$ beam splitters (BS). A
multichannel coincidence circuit effectively post-selects the terms of
the PDC state with one photon in each mode. As a result, with a
suitable choice of the relative phase between the photons of the
emitted pairs, we obtain correlations which characterize a
six-photon polarization entangled state given by the following
superposition of a six-qubit Greenberger-Horne-Zeilinger (GHZ) state
and two products of three-qubit W states:
\begin{equation}
\ket{\Psi_{6}^{+}} = \frac{1}{\sqrt{2}}\ket{\mathrm{GHZ}_{6}^{+}} +
\frac{1}{2}(\ket{\overline{W}_{3}}\ket{W_{3}}
+\ket{W_{3}}\ket{\overline{W}_{3}}), \label{state}
\end{equation}
where $\ket{\mathrm{GHZ}_{6}^{+}}=(\ket{HHHVVV}+\ket{VVVHHH})/\sqrt{2}$, and
$\ket{W_{3}}=(\ket{HHV}+\ket{HVH}+\ket{VHH})/\sqrt{3}$.
$\ket{\overline{W}}$ is the spin-flipped  $\ket{W}$, and $H$ and $V$
denote horizontal and vertical polarization, respectively.
\begin{figure}
\begin{center}
\includegraphics[width=8.5cm]{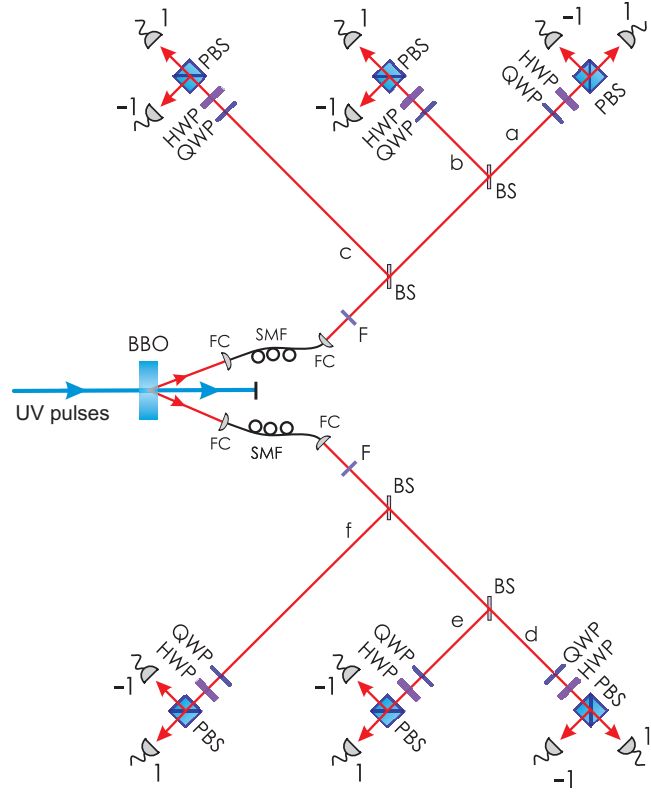}
\caption{\label{setup}\textbf{Experimental setup for generating and
analyzing the six-photon polarization-entangled state $\Psi_6^+$.}
The six photons are created in third order PDC processes in a 2 mm
thick BBO crystal pumped by UV pulses. The intersections of the two
cones obtained in non-collinear type-II PDC are coupled to single-mode 
fibers (SMFs) wound in polarization controllers. Narrow-band
interference filters (F) ($\Delta\lambda =3$ nm) serve to remove
spectral distinguishability between different signal-idler pairs.
The two spatial modes are  divided into three modes each by a
sequence of two $50-50$ beam splitters (BS). Each mode can be
analyzed in arbitrary polarization basis using half- and quarter
wave plates
   (HWP and QWP) and a polarizing beam splitter (PBS). Simultaneous detection
   of six photons (there is one detector at each output mode of the six polarizers) are recorded
   by a 12 channel coincidence counter.}
   \end{center}
\end{figure}

We have used a well tested setup of our laboratory, see~\cite{MOHAMED1}. 
A frequency-doubled Ti:Sapphire laser ($80$ MHz
repetition rate, $140$ fs pulse length), yielding UV pulses with a
central wavelength at $390$ nm and an average power of $1300$ mW, is
used as a pump. The laser beam is focused to a $160$ $\mu$m waist in
a $2$ mm thick BBO ($\beta$-barium borate) crystal. Half wave plates
and two $1$ mm thick BBO crystals are used for compensation of
longitudinal and transversal walk-offs. The  emission of
non-collinear type-II PDC processes is coupled to single-mode fibers
(SMFs). They collect radiation at the two spatial modes which are at
the crossings of the two frequency degenerated down-conversion
cones. After leaving the fibers the down-conversion light passes
narrow-band  ($\Delta\lambda =3$ nm) interference filters (F) and is
split into six spatial modes $(a, b, c, d, e, f)$ by ordinary
$50-50$ beam splitters (BSs), followed by birefringent optics to
compensate phase shifts in the BSs. Due to the short pulses, 
narrow-band filters, and single-mode fibers the down-converted photons are
temporally, spectrally, and spatially indistinguishable~\cite{ZZW95}, 
see figure~\ref{setup}. The polarization is being kept
by passive fiber polarization controllers. Polarization analysis is
implemented by a half wave plate (HWP), a quarter wave plate (QWP),
and a polarizing beam splitter (PBS) in each of the six spatial
modes. The outputs of the PBSs are lead to single-photon silicon
avalanche photodiodes (APDs) through multi-mode fibers. The APDs'
electronic responses, following photo detections, are being counted
by a multichannel coincidence counter with a $3.3$ ns time window.
The coincidence counter registers any coincidence event between the
$12$ APDs as well as single detection events.
\section{Analysis and Results}
\label{sec:results}
\subsection{The six-photon state}
\begin{figure}
\begin{center}
\includegraphics[width = 0.46\columnwidth]{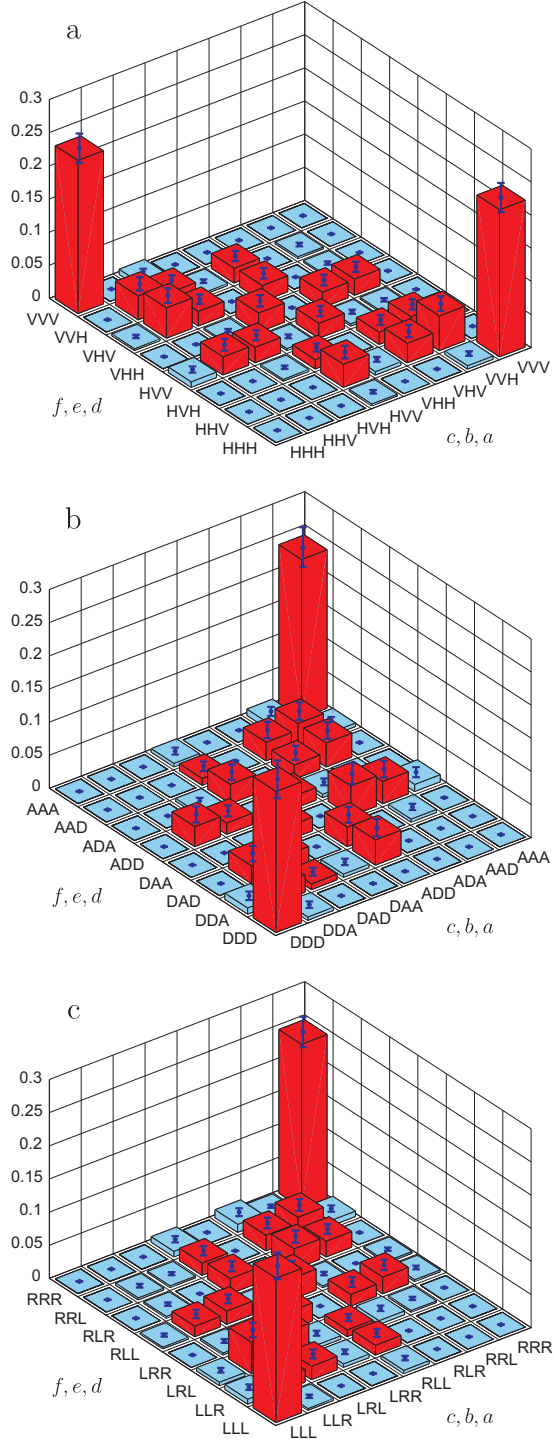}
\caption{\label{data}\textbf{Experimental results.} Six-fold
coincidence probabilities corresponding to detections of one photon
in each mode in the $\{\ket {H}, \ket{V}\}$ basis (a), $\{\ket{D},\ket{A}\}$ basis (b), and
$\{\ket{L},\ket{R}\}$ basis (c).
  The values of the correlation functions are $-89.5\%\pm4.9\%$,
   $+86.3\%\pm6.6\%$, and $+82.0\%\pm4.8\%$ respectively.
    For a pure $\Psi_{6}^{+}$
   state the light blue bars would be zero. In our experiment
    these values are all in the order of the noise.
    The measurement time was about 94 hours for each setting and the average six-photon detection rate was 3.4 events/hour.}
    \end{center}
\end{figure}

Figure~\ref{data}(a) shows experimentally estimated probabilities to
obtain each of the $64$ possible sixfold coincidences with one
photon detection in each spatial mode, for the case when all qubits
were measured in $\{\ket{H},\ket{V}\}$ basis. The peaks are in very
good agreement with theory:  half of the detected sixfold
coincidences are to be found as $HHHVVV$ and $VVVHHH$, and the other
half should be evenly distributed among the remaining events with
three $H$ and three $V$ detections. This is a clear effect of the
bosonic interference (stimulated emission) in the BBO crystal giving
higher probabilities for emission of indistinguishable photons.

The detection probabilities for our six-photon state reveal similar structure in the three
measurement bases $\{\ket {H}, \ket{V}\}$, $\{\ket{D},\ket{A}\}$
(diagonal/antidiagonal, $\ket{D/A} = (\ket{H} \pm
\ket{V})/\sqrt{2}$) and $\{\ket{L},\ket{R}\}$ (left/right circular,
$\ket{L/R} = (\ket{H} \pm i\ket{V})/\sqrt{2}$). 
Note that the structure would be exactly the same if the two swaps $D \leftrightarrow A$ (in figure~\ref{data}(b)) and $L \leftrightarrow R$ (in figure~\ref{data}(c)) were made in modes $a$, $b$ and $c$ or in $d$, $e$ and $f$. This corresponds to  adding a phase shift of $\pi$ between $H$ and $V$ in one of the two sets of modes.
The ideal state, $\Psi_{6}^{+}$, is invariant under identical unitary
transformations applied to each qubit, which leave the $\{\ket {H},
\ket{V}\}$ basis unchanged, but rotate the complementary ones.
Experimentally this can be revealed by using specific sets of {\em
identical} settings of all polarization analyzers. The results
should be similar for such settings. Our results for measurements in
diagonal/antidiagonal, and left/right circular polarization bases
are presented in figure~\ref{data}(b) and~\ref{data}(c). We clearly
observe the expected pattern, with a small noise contribution.
Moreover, the quiet uniform noise distribution in the three mutually unbiased measurement bases, makes it plausible to believe that the noise is close to white. Using this approximation we can estimate the effectively observed state as
\begin{equation}
	\rho_{exp} = p\ket{\Psi_6^+}\bra{\Psi_6^+} + (1-p)\eins^{\otimes 6}/2^6, \qquad 0\le p \le 1.
	\label{rho}
\end{equation}

\subsection{Five-photon states from projective qubit measurements}
\begin{figure}
\includegraphics[width = \columnwidth]{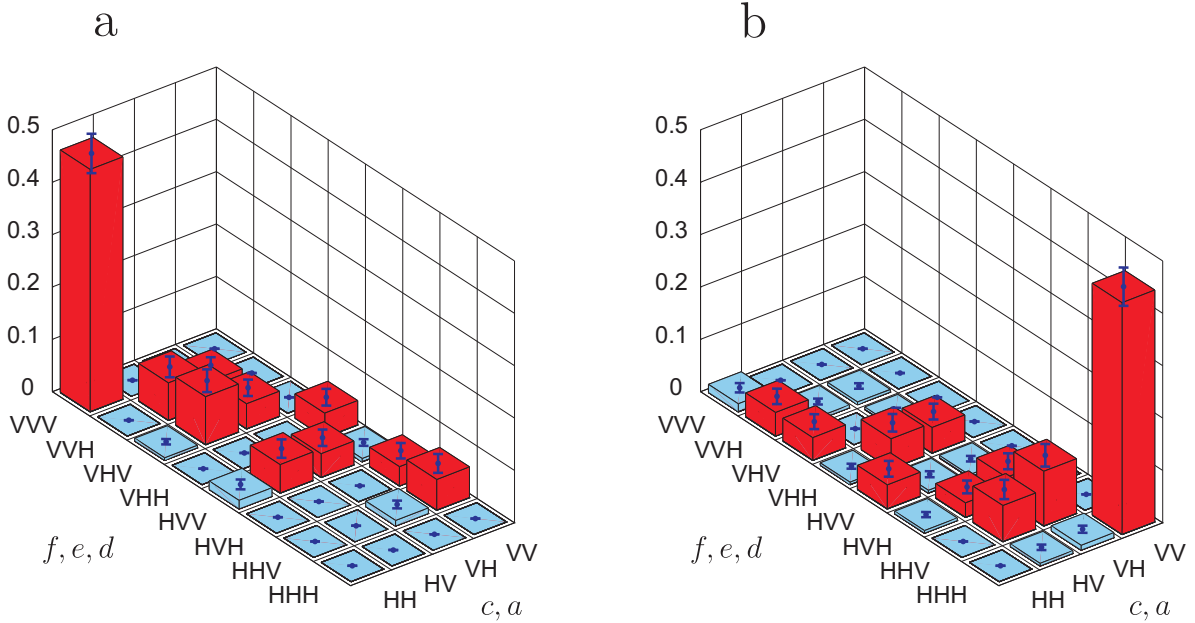}
\caption{\label{data5}\textbf{Five-photon states from projective
measurements.} Five-fold coincidence probabilities obtained through
the projection of the $b$-qubit (the photon in mode $b$) of $\ket{\Psi_{6}^{+}}$ onto $\ket{H}$ (a) and $\ket{V}$ (b), respectively. 
All qubits are measured in the $\{\ket {H}, \ket{V}\}$ basis.}
\end{figure}

The setup can also be  used to produce various five-photon states.
Conditioning on a detection of  one photon in a specific state we
obtain specific five-photon entangled states. In the computational
basis the projection of the second qubit onto $\ket{H}$ leads to
\begin{eqnarray}
_{b}\langle H \mid \Psi_{6}^{+}\rangle
=\frac{1}{\sqrt{2}}\ket{HHVVV}+\frac{1}{\sqrt{3}}
\ket{\Psi_{2}^{+}}\ket{\overline{W}_{3}}+\frac{1}{\sqrt{6}}\ket{VV}\ket{W_{3}},
\label{state5V}
\end{eqnarray}
while a  projection onto $\ket{V}$ results in
\begin{eqnarray}
_{b}\langle V \mid
\Psi_{6}^{+}\rangle=\frac{1}{\sqrt{2}}\ket{VVHHH}+\frac{1}{\sqrt{3}}
\ket{\Psi_{2}^{+}}\ket{W_{3}}+\frac{1}{\sqrt{6}}\ket{HH}\ket{\overline{W}_3}.
\label{state5H}
\end{eqnarray}
Figure~\ref{data5}(a) and~\ref{data5}(b) show the results related to these five-photon
conditional polarization states and  we clearly see the terms
$\ket{HHVVV}$ and $\ket{VVHHH}$, respectively. All these results are
in close agreement with theoretical predictions (up to the noise).

\subsection{Quantum correlation and entanglement}
Another property of $\ket{\Psi_{6}^{+}}$ is that, for certain
settings, it exhibits perfect six-qubit correlations. 
The correlation function is defined as the expectation value of the product of six local polarization observables.
Experimentally we have obtained the
following values: $\langle\sigma_z^{\otimes 6}\rangle = - 0.895 \pm
0.049$, $\langle\sigma_x^{\otimes 6}\rangle = + 0.863 \pm 0.066$
and $\langle\sigma_y^{\otimes 6}\rangle = + 0.820 \pm 0.048$, which 
are close to the theoretical values, $-1$, $+1$ and $+1$, respectively.
One can use these results to estimate $p$ from (\ref{rho}), as the average over the absolute values of the three correlations presented above.
To test the approximation (\ref{rho}) with the estimated value of $p=0.859\pm0.032$ we have also calculated the noise correlations in the three bases and obtained $\langle\sigma_z^{\otimes 6}\rangle_{\mathrm{noise}} = -0.035 \pm 0.051$, $\langle\sigma_x^{\otimes 6}\rangle_{\mathrm{noise}} = +0.004 \pm 0.067$
and $\langle\sigma_y^{\otimes 6}\rangle_{\mathrm{noise}} = -0.039 \pm 0.049$, which are all close to zero as is expected for white noise.
A rough measure of the fidelity can now be obtained through $F =
\langle\Psi_{6}^{+}|\rho_{exp}|\Psi_{6}^{+}\rangle = 0.861 \pm
0.031$. This is well beyond the results of other recent six-photon
experiments. 

$\ket{\Psi_{6}^{+}}$ is a genuine six-qubit entangled
state, meaning that each of its qubits is entangled with all the
remaining ones. In order to show that our experimental correlations
reveal six-qubit entanglement we use the entanglement witness method. An
entanglement witness is an observable yielding a negative value only
for entangled states, the most common being the maximum overlap
witness ($\WW_{max}$), which is the best witness with respect to
noise tolerance~\cite{witness}.
The maximum overlap witness optimized for $\ket{\Psi_{6}^{+}}$ has
the form
\begin{equation}
\WW_{max}=\frac{2}{3}\eins^{\otimes6}-\ket{\Psi_{6}^{+}}\bra{\Psi_{6}^{+}},
\end{equation}
where the factor 2/3 is the maximum overlap of $\ket{\Psi_{6}^{+}}$
with any biseparable state~\cite{TG05,Toth-MATLAB}. This witness
detects genuine sixpartite entanglement with a noise tolerance around $34\%$, but it
also demands a large number (183) of measurement settings. Since it would
be an experimentally very demanding task to perform all these
measurements, we have developed a reduced witness that can be
implemented using only three measurement settings. Our reduced
witness $\WW$, is given by
\begin{eqnarray}
\mathcal{W}=&\frac{181}{576}\eins^{\otimes6}-\frac{1}{64}(\sigma_{x}^{\otimes6}+\sigma_{y}^{\otimes6}-\sigma_{z}^{\otimes6})
-\frac{1}{576}\sum_{i=x,y,z}\big(3\sigma_{i}^{\otimes2}
\eins^{\otimes4}+3\sigma_{i}\eins\sigma_{i}\eins^{\otimes3} \nonumber \\
&
+3\eins\sigma_{i}^{\otimes2}\eins^{\otimes3}+3\eins^{\otimes3}\sigma_{i}^{\otimes2}
\eins+5\sigma_{i}^{\otimes2}\eins\sigma_{i}^{\otimes2}\eins+5\sigma_{i}\eins\sigma_{i}^{\otimes3}\eins
 \nonumber \\
&
+5\eins\sigma_{i}^{\otimes4}\eins+3\eins^{\otimes3}\sigma_{i}\eins\sigma_{i}+5\sigma_{i}^{\otimes2}
\eins\sigma_{i}\eins\sigma_{i}+5\sigma_{i}\eins\sigma_{i}^{\otimes2}\eins\sigma_{i}
\nonumber \\
&
+5\eins\sigma_{i}^{\otimes3}\eins\sigma_{i}+3\eins^{\otimes4}\sigma_{i}^{\otimes2}+5\sigma_{i}
^{\otimes2}\eins^{\otimes2}\sigma_{i}^{\otimes2}+5\sigma_{i}\eins\sigma_{i}\eins\sigma_{i}^{\otimes2}
 \nonumber \\
&
+5\eins\sigma_{i}^{\otimes2}\eins\sigma_{i}^{\otimes2}+[\eins\leftrightarrow\sigma_{i}]_{i=x,y}-[\eins\leftrightarrow\sigma_{z}]
\big),
\label{witness}
\end{eqnarray}
where $[\eins\leftrightarrow\sigma_{i}]$ denotes 
the same terms as in the sum but with $\eins$ and $\sigma_{i}$ interchanged. 
It is obtained
from the maximum overlap witness as follows. First the maximum
overlap witness is decomposed into direct products of Pauli and
identity matrices, next only terms that are tensor products of
$\sigma_i$ with a fixed $i$ and of identity matrices are selected
(all terms that include products of at least two different Pauli
matrices are deleted). Finally, the constant in front of
$\eins^{\otimes6}$ in the first term of (\ref{witness}) is
chosen to be the smallest possible such that all entangled states that are
found by the reduced witness are also found by the maximum overlap
witness. Our reduced witness detects genuine sixpartite entanglement
of $\ket{\Psi_{6}^{+}}$ with a noise tolerance of $15\%$. The
theoretical expectation value $\langle\WW_{th}\rangle=-1/18
\approx-0.056$ and our experimental result is
$\langle\WW\rangle=-0.021\pm0.014$, showing entanglement with an
accuracy of $1.5$ standard deviations.

Furthermore, the data that we have acquired allows one to use the so-called entanglement indicator method 
proposed in~\cite{GEOMETRIC} to verify entanglement in the observed correlations. This method is based on comparisons of scalar products of correlation tensors of separable states and the state that one tests for entanglement. Here we shall present a {\em modified} version of this method based on norms. Any fully separable six-qubit state has six-qubit correlations, each of which are described by a convex combination of a set of tensor products of Bloch vectors describing pure state qubits. That is, the six-qubit correlation tensor $T_{i_1 \dots i_6}=\langle\sigma_{i_1}\otimes \dots \otimes\sigma_{i_6}\rangle$, where $i_k=x,y,z$, of a fully separable state is given by a convex combination of $T_{pure}=\vec{t}_1\otimes\dots\otimes\vec{t}_6$, where $\vec{t}_i$ are normalized three-dimensional (Bloch) vectors. Since the norm of $T_{pure}$, treated as a $3^6$-dimensional vector, is $1$, any convex combination of such tensors has a norm which is maximally one. This is a generic property of normalized vectors in any space. Consequently, if the norm of the correlation tensor of the tested state, that is $\sum_{i_1, \dots ,i_6}T^2_{i_1 \dots i_6}$, is greater than one, the state cannot be fully separable. Clearly, if any partial sum of squares of the correlation tensor elements exceeds $1$, the same conclusion is valid.  
From our measurement data we obtain that $\langle\sigma_z^{\otimes 6}\rangle^2+
\langle\sigma_x^{\otimes 6}\rangle^2+ \langle\sigma_y^{\otimes
6}\rangle^2$ has the value of $2.22 \pm 0.16$, which is much
greater than $1$ (by $7.4$ standard deviations). Note that even if we sum only two of these three components, we get around $1.48$ and the entanglement is revealed. This clearly demonstarates the ``friendliness'' of the method, as well as the strength of the observed entanglement. 

\section{Conclusion}
\label{sec:conclusion}
In summary, we have experimentally demonstrated that six-photon
correlations specific for the $\Psi_6^+$ state are experimentally
observable. This is done with a previously tested setup~\cite{MOHAMED1}, which uses a suitable
filtering/selection procedure to single out triple emissions from a {\em single} pulsed PDC source. 
We have analyzed the six-qubit state in three measurement bases and our six-photon coincidences follow the interference characteristics for $\ket{\Psi_{6}^{+}}$. Moreover, the noise contribution in our experiment is quite low and the collected data are of a high fidelity with respect to theoretical predictions. 
We have used the entanglement
witness method to detect sixpartite entanglement in the state, as well as introduced a new version of the indicator method to reveal entanglement in the data.
The high fidelity of the observed state and the high stability of
our interferometric-overlap-free setup makes the six-photon source
useful for multiparty quantum communication and particularly for the
demonstration of the telecloning communication scheme.
For the implementation of three-location-telecloning, we will use a brighter source and similar multiphoton interference techniques as are reported in this work.

\ack This work was supported by the Swedish Research Council (VR).
M.\.{Z}. was supported by the Wenner-Gren Foundations and by the EU
program QAP (Qubit Applications, Grant No. 015858).

\end{document}